# On the Coupling of Photon Spin to Electron Orbital Angular Momentum

U.C.Fischer[1*], F. Fontein[1], H. Fuchs[1], R. Salut[2], Y. Lefier[2], T. Grosjean[2]

[1]Interface Physics group, Departnment of Physics, Westfaelische Wilhelms-Universitaet, Muenster, Germany

[2] Department of Optics, FEMTO-ST Institute, UMR CNRS 6174, University of Franche-Comté, Besancon, France

E-mail: fischeu@wwu.de

**Abstract**  Partially gold coated 90° glass wedges and a semi - infinite slit in a thin film of gold ending in a conducting nano-junction serve as samples to investigate the transfer of photon spin to electron orbital angular momentum. These structures were specifically designed as samples where an incident beam of light is retroreflected. Since in the process of retroreflection the turning sense of a circularly polarized beam of light does not change and the direction of propagation is inverted, the photon spin is inverted. Due to conservation of angular momentum a transfer of photon spin to electron orbital angular momentum of conduction electrons occurs. In the structures a circular movement of electrons is blocked and therefore the transfered spin can be detected as a photovoltage due to an electromotive force which is induced by the transfer of angular momentum. Depending on the polarization of the incident beam, a maximum photovoltage of about 0,2µV was measured for both structures. The results are interpreted in terms of a classical electrodynamic model of the monochromatic linearly polarized photon as a propagating solitary electromagnetic wave of finite energy h$\nu$ which carries an angular momentum $\frac{h}{2\pi}$ which is elaborated elsewhere where $h$ is Planck's constant and $\nu$ the frequency of light. The relative values of the measured photovoltages for different polarizations can well be explained by the electrodynamic model of a photon and an associated spin angular momentum. The absolute values of the measured photovoltages are also consistent with the interpretation. The observed effects are closely related to the lateral Fedorov Imbert shift of focused beams in optics and the optical spin Hall effect and to other non linear optical effects such as the inverse faraday effect for which a new interpretation is given here in terms of the electrodynamic model of the photon and its spin.

**Introduction**

Light and other electromagnetic radiation carries linear and angular momentum and therefore not only energy but also linear and angular momentum can be exchanged between electromagnetic radiation and matter. Longitudinal and transversal photoinduced electrical currents were previously measured on metal films and periodically structured metal films [1-5]. Different explanations were given for the observed effects in terms of momentum transfer as well as transfer of spin and angular momentum from electromagnetic radiation to electron movement and of the rectification of electromagnetic fields. We fabricated metal nanostructures which are connected to macroscopic electrodes in order to investigate the photoinduced generation of an electromotive force on a single metal nanostructure by measuring the photovoltage generated between the electrodes. Our specific aim was to detect photovoltages due to the transfer of photon spin to electron orbital angular momentum.

Mansuripur et al [6] addressed theoretically the problem of transfer of spin angular momentum to orbital angular momentum of a normally incident beam of circularly polarized light into an ideally conducting concave 90° wedge when this beam of light is reflected from this structure. A normally incident beam is in this case not specularly reflected as from a normal mirror but it is reflected into the reversed direction of its incidence. An interesting property of retroreflectors is, that for circularly polarized incidence, the reflected beam has the same turning sense as the incident beam whereas for specular reflection from a planar mirror, the turning sense is reversed. If the turning sense of the beam remains the same after an inversion of the direction of propagation, the spin angular momentum S of the light beam has to be inverted in the process of the retroreflection and the reflected beam carries an angular momentum -S. Because of conservation of angular momentum an angular momentum L = 2S has to be transferred to the mirror in the process of retroreflection. If the spin is transferred to electron orbital angular momentum, spin transfer could be detected by measuring light induced currents or an associated magnetic field. We chose a different way. Instead of measuring light induced currents we measure a light induced electromotive force in structures where the circular electron movement is blocked.

A linearly polarized beam of light can be considered as a current of photons of electric radiation originating from electric dipole transitions [7]. These photons are Bosons with a spin 1 carry an angular momentum $J = h/2\pi$ [8], where h is Planck's constant. As we want to address the question of transfer of photon spin to electron orbital angular momentum we do not only consider the transfer of angular momentum from circularly polarized light but also from linearly polarized light.

For an interpretation of the results we use a classical electrodynamic model of a monochromatic propagating photon of a frequency $\nu$ as a solitary electromagnetic wave of a finite energy content $h\nu$ and a spin angular momentum $\frac{h}{2\pi}$, which is oriented in the direction of polarization and the longitudinal propagation direction as indicated in fig. 1.

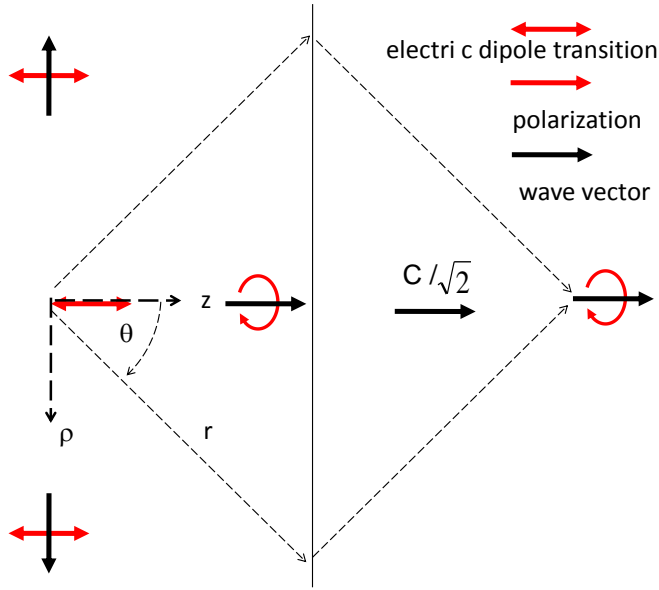

**Fig. 1** *Schematic view of a photon emitted by an electric dipole transition propagating in the longitudinal direction, which is focused by means of a lens. The center of field energy of the photon propagates in the longitudinal direction at a speed of $\frac{c}{\sqrt{2}}$..The radiation in the longitudinal direction is right handed circularly polarized, whereas the emission in the transverse diections is linearly polarized.*

This model is described in detail elsewhere [9]. It is strongly inspired by the concept of P.A.M. Dirac of a polarized photon originating from electric dipole radiation [10]. Dirac defines the photon as electromagnetic radiation which is emitted in an electric dipole transition. The dipole emission of positive frequency in a longitudinal direction parallel to the orientation of the dipole has a right handed circular polarization (left handed circular polarization corresponds to a negative frequency). The concept of a longitudinally propagating photon may at first sight be confusing since the radiation of the dipole is known to be directed perpendicular to the orientation of the dipole. However a longitudinal propagation of dipole radiation can easily be detected e.g. by focusing the longitudinally emitted radiation with a lens as indicated in the fig.1. There it is also seen, that the photon propagates in the longitudinal direction at a speed of $\frac{c}{\sqrt{2}}$ and not at the speed of light. In the direction perpendicular to the dipole the radiation is linearly polarized and in other directions the light has an elliptic polarization. In our model the spin of the photon can be described as a property of the classical electromagnetic radiation field of a propagating monochromatic wave of energy content hν and angular momentum $\frac{h}{2\pi}$. Within this model the photon has no oscillatory electromagnetic field but it consists of a single cycle solitary wavepacket propagating at a speed of $c/\sqrt{2}$ in the longitudinal positive z direction of the dipole orientation. These properties of our model of a photon are sufficient for an interpretation of the experimental results reported here. We point out that our model of the photon is a non linear model. For an interpretation of the transfer of photon spin to electron orbital angular momentum we need a non linear model of the time harmonic monochromatic photon, because the scattering process leading to the transfer of angular momentum is by necessity an inelastic process where not only angular momentum is transferred from the photon but also kinetic energy which is associated with an angular momentum of electrons. The time harmonic property of the linearly polarized photon in our model is not encoded in an oscillation of the electromagnetic field of a stationary planar wave at a fixed frequency but in the angular frequency associated with its spin angular momentum. The concept is thus closely related to the concept of a rotating arrow for an elementary particle described by Feynman [11]. This angular frequency decreases in the elementary scattering process of the photon which undergoes a reflection in a retroreflecting structure. In our interpretation of the experimental results, evanescent waveguide and edge plasmon modes play an important role. These modes are mixed longitudinal and transversal modes which carry an angular momentum. Their importance in the interpretation of the transfer of angular momentum from electromagnetic circular polarized radiation to matter was investigated previously [12,13]

**Experimental details**

We devised 2 different structures for the measurements.

1.The metal coated finite glass wedge. A metal coated tetrahedral tip of glass was introduced previously as a probe for Scanning Near-Field Optical Micoscopy [14]. This dielectric tip has the property of total retroreflection and we therefore use a similar tip for an investigation of the transfer of photon spin to electron orbital angular momentum. The tip is used in 2 different optical configurations as shown schematically in Fig.2a,b ,which we call a) the tip configuration and b) the wedge configuration respectively. In the tip configuration a focused beam of light is incident into the tip at an angle of 45 ° with respect to a metal coated edge. In the wedge configuration, the beam is incident normally into the wedge. A wedge is formed by breaking a piece of cover glass of a thickness of 0,18 mm. A 90° wedge is thus formed between the smooth surface of the glass and the fracture plane with a common fracture edge of a radius of curvature in the order of 1 - 5 nm. Both faces of the wedge are coated with a 20 nm thick film of gold by separate metal deposition steps at an oblique angle with respect to the faces of the wedge such that the edge is coated with a thinner metal film by controlling the evaporation angle which leads to a decreased metal deposition at convex edges. In a subsequent step the glass is broken again at an angle of 90° with respect to the wedge. In this way a tetrahedral glass tip is formed, where 2 faces of the tip are coated with metal. In a further step, the metal coating of the edge is partially removed mechanically such that the metal coating of the edge extends to only about 30 µm away from the tip for the tip configuration. For the wedge configuration the metal coating of a 5 mm wide wedge was removed mechanically at both sides of the wedge leaving a metal coated edge of a length of about 100 µm at the center of the wedge. In order to be able to illuminate the edge or the tip with a beam of light, the glass fragments were glued onto a 90° prism by a transparent index matched glue. For electrical measurements the 2 metal faces of the wedge were connected to separate electrodes. For the tip configuration, the metal coating on the edge, which was confined to a length of about 30 µm, formed the metal junction between the 2 gold coated faces of the wedge with a resistance of 340 $\Omega$. In the case of the wedge configuration the central metal coated edge formed a junction with a resistance of 12 $\Omega$ between the 2 faces of the wedge. The thickness of the gold film on the edge must have been much smaller than 20 nm, because the resistance across the edge was of the same order of magnitude as the resistance of the very narrow gold junction of a cross section in the order of 400 nm$^2$ in the case of the semiinfinite slit sample described below and we assume that the metal junction across the edge consisted of a discontinuous metal film with only a few electrically conducting junctions between the two faces of the wedge.

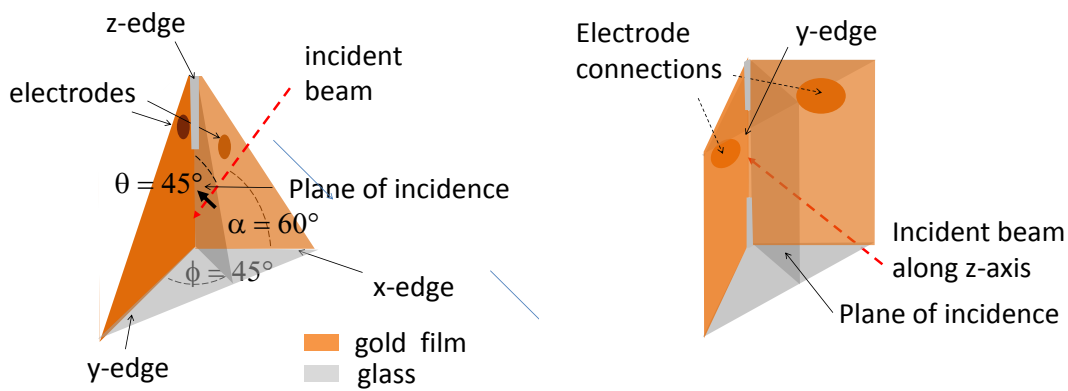

**Fig. 2 a**                          **b**

*Schematic view of experimental scheme for finite wedge samples a) in a tip configuration. For a p-polarized beam incident into the z-edge the black fat arrow indicates the longitudinal orientation of the photon spin in the direction of polarization b) in the wedge configurateion.*

**2. Semiinfinite Slit in a thin metal film.** As a second structure we use a narrow slit in a 20 nm thick film of gold which was deposited on a 180µm thick cover glass. By using a focused Ion beam a long narrow slit was cut into the gold film. A wide slit was cut perpendicular to the narrow slit, such that at the end of the slit a narrow metal junction was formed between the two sides of the gold film separated by the slit. Several examples of such slits, which were used for measurements are shown in fig 3. The experimental configuration for the semiinfinite slit is shown schematically in fig. 4.

Setup for photoelectrical measurements. For illumination we used 0,5 mW 655 nm diode laser. The laser beam was modulated with a 1 kHz signal from a function generator. The polarization of the light beam was adjusted by a LC crystal plate. The beam was focused with a microscope objective lens with a numerical aperture of approximately 0,1 into the different samples. For the tip configuration the beam was carefully focused into the apex ot the tip and the beam was not displaced from this position in the course of measurements. For the wedge configuration the beam of light was incident normal to the edge of the wedge. For the wedge

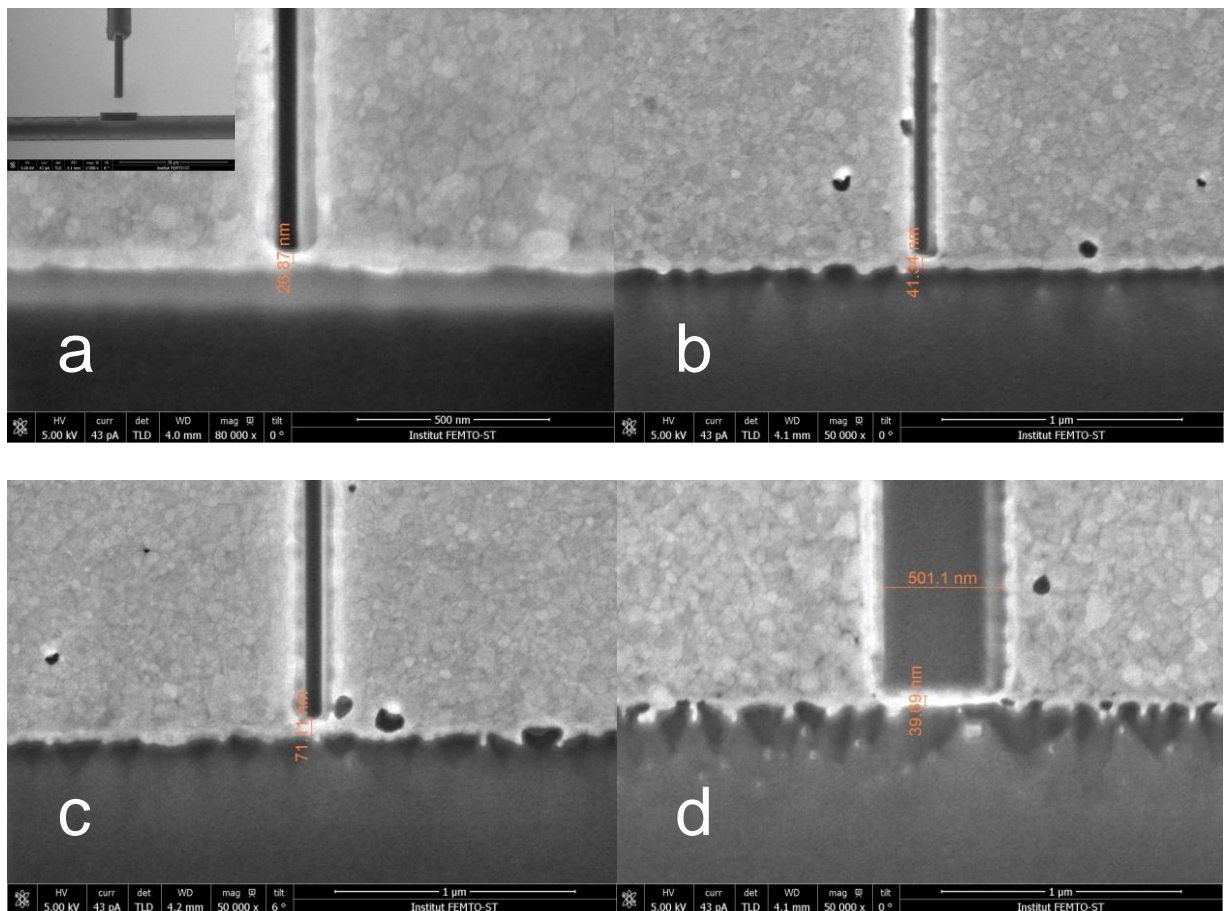

**Fig. 3**

*Electron Micrographs of "semiinfinite slit " samples. a) sample 1: slit width about 50 nm and width of metal junction 27 nm; the resistance of the metal junction was 15 Ω. Inset: Overview. A thin slit is cut perpendicular to a wide slit into a 20 nm thick gold film with a 1 nm chromium adhesion layer on a glass cover plate. The thin slit is continued by a coarse slit in order to electrically separate the 2 adjacent sides of the slit. b) sample 2: slit width about 50*

nm and width of metal connection 41 nm, resistance of metal junction 12 Ω. c) sample 3: width of metal connection 73 nm, resistance of metal junction 3 Ω.d) sample 4; slit width about 500 nm and width of metal junction 40 nm; the resistance of the metal junction was 60 Ω.

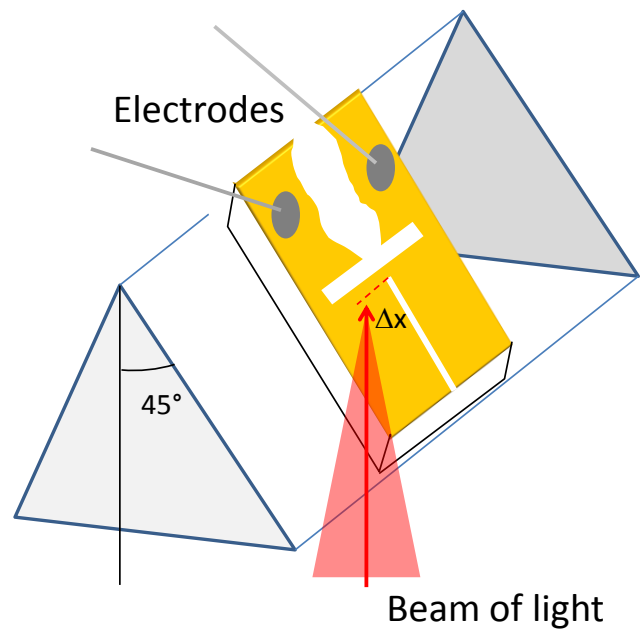

**Fig. 4**

*Schematic view of experimental scheme for "semiinfinite" slit samples. The sample is optically connected to a prism. The end of the slit is irradiated with a focused beam of light of a wavelength of 655 nm, an intensity of 0,5 mW and a numerical aperture of 0,1. A Kretschmann configuration is used with an angle of incidence of 45°. The 2 sides of the metal film separated by the slit are connected to electrodes. The photovoltages are measured as a function of the polarization and lateral displacement Δx of the incident beam.*

and slit samples the beam could be mechanically displaced in an axis perpendicular to the slit or parallel to the wedge respectively and the displacement was recorded with an accuracy of 1 µm. The voltage between the electrodes of the sample was fed into a lock in amplifier and the 1 kHz in phase AC signal was recorded with an integration time of 50s and a settling time of 300 s to reach a stationary value. The sensitivity was limited by a background signal in the absence of illumination in the order of +/- 10 nV. A 90° out of phase signal was also observed but variations due to irradiation were limited to +/- 20 nV, very close to the detection limit. No further attention was yet paid to this 90° out of phase signal.

**Experimental Results**

<u>Wedge in the tip configuration.</u> The experimental results for the tetrahedral tip are shown in Table 1 for different polarizations of the incident beam.

| polarization | rcp | lcp | p | s |
|---|---|---|---|---|
| experimental value /µV | 0,195 +/- 0,01 | 0,107 +/- 0,01 | 0,157 +/- 0,01 | 0,147 +/- 0,01 |
| normalized experimental value | 1 | 0,55 | 0,805 | 0,754 |
| normalized calculated value | 1 | 0,57 | 0,81 | 0,81 |
| Calculated spin transfer/ $\frac{h}{2\pi}$ | $\sqrt{\cos^2(60°)+\cos^2(45°)}$ | $\cos(60°)$ | $\cos(45°)$ | $\cos(45°)$ |

**Table 1** *experimental results for different polarizations of the incident beam, right- left handed circular polarized (r-l cp), p- and s- polarized. Comparison to calculated values*

A signal of 0,2µV was obtained for right handed circular polarized light and of 0,1 µV for left handed circular polarized light. For s and p polarized light with the polarization in the plane of incidence or perpendicular to the plane of incidence respectively, a signal of 0,150 µV was obtained just half way in between the 2 circular polarizations.

<u>Wedge with normal incidence</u>

We first performed measurements with p-polarized incidence. The beam was focused onto center of the metal coated edge of the wedge and measurements were performed for different positions of the beam when the beam was displaced along the x axis normal to the edge. The results of these measurements are shown in fig.5 In this case a maximal voltage of 0,14. µV was obtained with a half width of 4 µm. After a new optimized adjustment of the beam, measurements for different polarizations were recorded and the results are shown in table 2. The largest signal of 0,194 µV was obtained for p polarization and the smallest signal of 0,162 µV was obtained for s-polarization. Within the accuracy of the measurements the same signal of 0,180 µV was obtained for left and right hand circular polarized light just half way between s- and p- polarized light. The values for the different polarizations are quite different for the tip and wedge configurations respectively but the maximum signals happened to be almost the same for the tip and the wedge configurations.

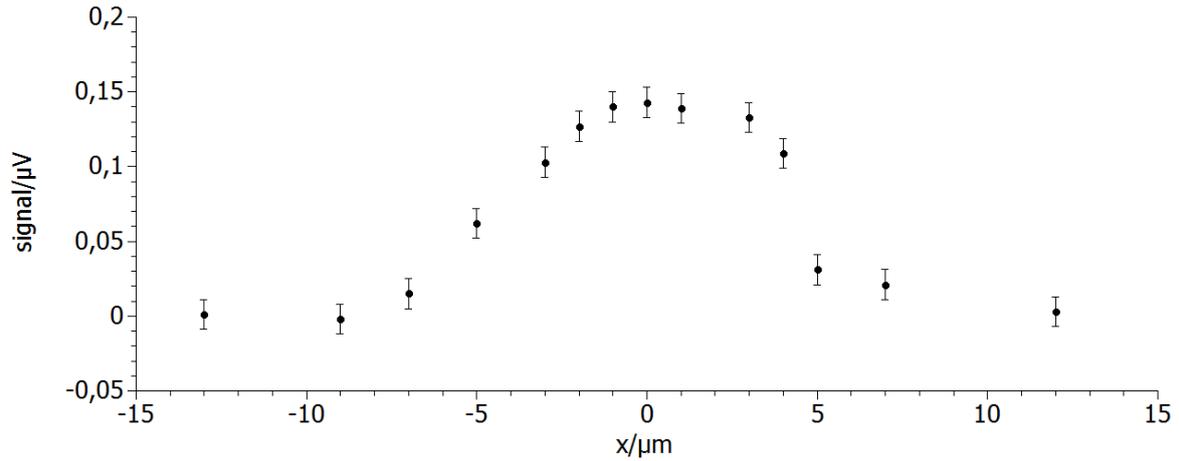

**Fig. 5** *Experimental results for the wedge configuration with normal incidence of p-polarized light as a function of lateral displacement x of the beam.*

| polarization | p-y | s-x | lcp | rcp | no light |
|---|---|---|---|---|---|
| signal /µV ± 0,01 | 0,194 | 0,162 | 0,178 | 0,182 | 0,001 |
| normalized signal ± 0,05 | 1 | 0,83 | 0,92 | 0,94 | 0,005 |
| Normalized spin transfer | 1 | 0,87 | 0,935 | 0,935 | |
| Calculated spin transfer / $\frac{h}{\pi}$ | 1 | $\sqrt{cos^2 45° + cos^2 60°}$ | $\frac{1 + \sqrt{cos^2 45° + cos^2 60°}}{2}$ | | |

**Table 2** *experimental results for the wedge configuration with normal incidence for different polarizations of the incident beam. p-y signifies p-polarization along the y-axis, s-x signifies s polarization along the x axis, lcp and rcp signify left and right hand circular polarization respectively.*

<u>The semiinfinite slit.</u>

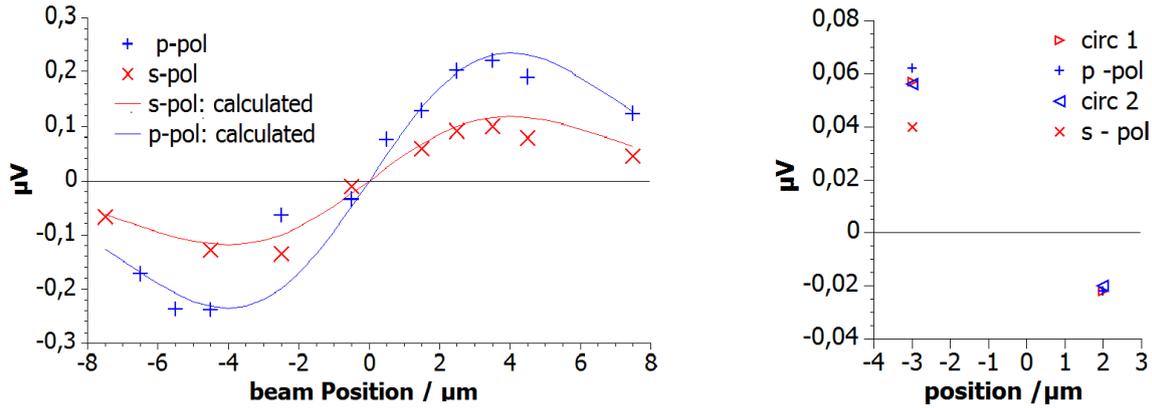

**Fig. 6 a** *experimental results for sample 1*  **b** *for sample 2*

Fig. 6 a shows the results for the semiinfinite slit sample 1. For all polarizations a signal was observed which changed sign when the position of the focused beam was changed from one side of the slit to the other side. For p polarization a maximum signal of +/- 0,23 µV was observed at a lateral displacement of the beam of about 4 µm from the center of the slit. whereas for s polarization the maximal signal of +/- 0,12 µV signal was only about half as big. Measurements for circular polarizations were not yet performed for this sample. For sample 2 less data were recorded, but the maximum signal was only about 50 nV as shown in fig. 6b. No significant signal was obtained for samples 3 and 4. In sample 1 the metal junction had a width of 27 nm, a length of 50 nm and a resistance of 15 Ω. In sample 2 the width of metal connection was 41 nm, and it had a resistance of 12 Ω. In sample 3 the width of metal connection was 73 nm, and it had a resistance of 3 Ω. In sample 4 the slit width was about 500 nm and the width of the metal junction was 40 nm; the resistance of the metal junction was 60 Ω. In the case of sample 1 where we observed the strongest effect, the width of the metal junction of 27 nm is in the order of the skin depth of gold at 655 nm. Our results thus show that with increasing width of the junction decreases significantly and also an increased length and resistance of the junction decreases the effect.

**Interpretation of experimental results**

In order to interpret the observed photovoltages we consider an incident Gaussian beam as a current of photons which interacts with the metal structures. According to Huygens principle, a propagating beam of light is composed of elementary wavelets. In terms of our model of a photon we consider the wavelets to be photons of the same polarization as the incident beam of light. Let us first consider the wedge in the tip configuration. The photon spin movement in the process of retroreflection can be completely traced by considering how leaky edge – waveguide modes [15] excited at the x-y-z edges by the incident beam propagate towards the tip and are there transmitted to edge waveguide modes along the other edges which finally add up to the reflected beam propagating away from the tip. As indicated schematically in fig.2a, a p-polarized incident focused beam carries in terms of our model of a photon spin angular momentum $h/2\pi$ oriented in the direction of polarization at an angle of 45° with respect to the z- axis with a z component $\frac{h}{2\pi}\cos 45°$. The incident beam excites a leaky z- edge waveguide mode propagating towards the tip and this z-edge waveguide mode carries the same spin as the incident beam due to the conservation of angular momentum in the process of excitation of the edge waveguide mode. The z–edge waveguide mode is transmitted at the tip into x- and y- edge waveguide modes which propagate in the negative x- and y- direction and which have no z –oriented spin component, In the transmission process, the z oriented spin changes from $\frac{h}{2\pi}\cos 45°$ to 0. Due to conservation of angular momentum this change of spin has to be counterbalanced by a transfer of angular momentum from the photon to an angular momentum $-\frac{h}{2\pi}\cos 45°$ of the mirror. This angular momentum can consist in a circular movement of the mirror or in a circular movement of electrons. Due to the high mass of the mirror we assume that the momentum will exclusively be transferred to conduction electrons of the metal. The orbital angular momentum of the electron is induced by a torque of the photon which acts on the electron. This torque can be considered as an inertial torque of the photon due to the time derivative of the angular momentum of the spin of the photon when it changes its orientation. The same spin transfer occurs for an s – polarized incident beam, which excites an x- edge waveguide mode and a y- edge waveguide mode with a spin $\frac{h}{2\pi}\cos 45° \cos 60°$ each with a z component $\frac{h}{2\pi}\cos^2 45° \cos 60° = \frac{h}{8\pi}$ in the z –direction. The z-components add up at the tip to $\frac{h}{4\pi}$ and in the transmission of the x-y edge waveguide modes to a z- edge waveguide mode the z component of the spin flips from the negative to the positive z-direction. Again due to conservation of angular momentum, the spin has to be transferred to a counterbalancing electron orbital angular momentum $-\frac{h}{2\pi}\cos 45°$ in the z- direction .

We now consider the rcp incident beam. It excites a z – edge waveguide mode propagating towards the tip with a spin oriented in the direction of propagation of the incident beam. the z- edge waveguide mode has a z component of the spin of $\frac{h}{2\pi}\cos 45°$ . At the tip, the z-edge waveguide mode is transmitted to x- and y- edge waveguide modes propagating away from the tip, which carry no z oriented spin. But the rcp beam also excites edge waveguide modes along the x-and y- edges  propagating in the –x and –y directions. They carry a spin component of $\frac{h}{2\pi}\cos 60°$ in the direction perpendicular to z, as the angle α between  the incident beam and the x- and y- edges is in this configuration 60° (see fig. 6). The x- and y- edge waveguide modes are transmitted at the tip into a z-wedge waveguide mode. In the hole process of retroreflection of a rcp beam a spin of $\sqrt{\left(\frac{h}{2\pi}\cos(60°)\right)^2 + \left(\frac{h}{2\pi}\cos(45°)\right)^2} = 0{,}87\frac{h}{2\pi}$ is transferred to electron orbital angular momentum. It compensates the change of photon spin in the transmission of the z- edge waveguide mode to the x-y wedge waveguide modes and vice versa. On the other hand, in the case of a lcp incident beam only x- and y-wedge waveguide modes propagating towards the tip are generated, which lead to a transmission to z-wedge waveguide mode. In this transmission process a spin of $\frac{h}{2\pi}\cos(60°) = 0{,}5\frac{h}{2\pi}$ is transferred to electron orbital angular momentum.

Table 1 shows a comparison of the experimental results which were normalized to the maximum of the experimental values for the measured electromotive force and the calculated normalized values for the transfer of photon spin to electron angular momentum for the tip configuration of fig 3a. It is seen that our interpretation leads to a coincidence of these experimental and calculated values respectively.

Let us now turn to the case of normal incidence into the wedge of a p-polarized Gaussian beam. At first sight it is surprising that we can detect a photoinduced voltage difference between left and right because there is no obvious geometric asymmetry in the experimental configuration. For an interpretation of the results we have to consider in this case only edge waveguide modes along the y-edge. A p-polarized Gaussian beam normally incident into the wedge in the + z direction has longitudinal components propagating in the ±y directions carrying a spin oriented in the longitudinal direction of propagation which excite edge waveguide modes propagating in the ±y direction. In the same way as the edge waveguide modes are excited these leaky waveguide modes reemit into the negative z- direction. This reemission of the leaky waveguide mode generates the totally reflected beam. The incident photon, which carries a spin $S = \frac{h}{2\pi}$, excites longitudinal edge waveguide modes propagating in the $+y$ and $-y$ direction with right handed and left handed circular polarization respectively. Since these edge modes propagate in the opposite y- directions they have the same spin $S_y = \frac{h}{2\pi}$. In the reflection process the y oriented spin is inverted due to 2 consecutive reflections of the excentric x - components of the beam at the 2 sides of the edge and the reflected beam carries no spin in the y direction. Therefore, due to the spin flip and conservation of angular momentum an electron orbital angular momentum $L_y = \frac{h}{\pi}$ compensates the spin flip. We can thus understand our experimental result of a photovoltage. An s-polarised incident beam carries a spin oriented in the x-direction. A photon excentrically incident from the $\pm$x direction excites an edge waveguide mode propagating in the $\pm$y direction respectively carrying a y directed spin $S_y = \frac{h}{2\pi}\sqrt{cos^2 45° + cos^2 60°} = \frac{0,87\,h}{2\pi}$. Due to the 2 consecutive reflections of the excentric components in the process of the reflection again the spin is inverted in the process of total reflection in the wedge and due to the conservation of angular momentum an electron orbital momentum $L_y = \frac{0,87h}{\pi}$ compensates the spin flip. For both circular polarisations we expect the same intermediate value $L_y = \frac{0,94\,h}{\pi}$

The calculated normalised values of the electron orbital angular momentum thus coincide with the normalised values of the measured photovoltages shown in Table 2.

Let us now consider the case of slit structure. For an interpretation we consider edge plasmon excitations at the edges of the metal film. Edge plasmons play here a similar role as the wedge waveguide modes in the case of the tetrahedral tip. In the case of the tetrahedral tip, waveguide modes along the x, y, and z- axis are involved in the spin transfer process. In the case of the semiinfinite slit the situation is a bit different, but also 3 edges are involved. The 2 edges on either side of the slit along the y axis and the metal film edge along the x – axis perpendicular to the slit. The photovoltage is generated between the 2 sides of the slit as a result of the transfer of edge Plasmon modes from one edge to another edge if a change in the z-component of the associated spin occurs in the transfer process.

In order to interpret the observed photovoltages for the slit structure we consider an incident gaussian beam as a current of incoherent photons which interact independently with the metal structures. According to Huygens principle, a propagating beam of light is composed of elementary wavelets. In terms of our model of a photon we consider the wavelets to be photons of the same polarization as the incident beam of light. Depending on their polarization, the incident photons can excite edge plasmon modes at the 2 edges oriented along the slit structure and at the perpendicular edge at the end of the slit which is oriented along the x-axis. We consider the interaction of the photons with the slit structure in terms of an excitation of edge plasmon modes which propagate towards the junction at the end of the slit, where they are either reflected or transmitted to edge plasmon modes at other edges. We first consider the spin transfer for p-polarized incidence as shown in the scheme in fig.8a. For p-polarization, photons have a polarization in the plane of incidence of the incident beam and they carry a spin in the direction of polarization with a positive z – component $S_z = \frac{h}{2\pi}\cos 45°$. For an incidence of a photon from the left, an edge plasmon mode is excited at the left edge of the slit and at the x edge both propagating towards the junction. They both carry a z-component $S_z = \frac{h}{2\pi}\cos 45°$ of the longitudinal spin of the incident beam. The edge plasmons are transmitted at the constriction to the right edge of the slit and a reflected Plasmon with left handed circular polarization propagates in the –y direction. In this process the z spin component is inverted. The change in Spin angular momentum $\Delta S_z = -\frac{h}{\pi}\cos 45°$ at the right edge is compensated by an electron orbital angular momentum . For incidence of the p-polarized photon at the right side of the slit, an electron angular $L_z = -\frac{h}{\pi}\cos 45°$ is generated accordingly. For incidence of a photon along the y axis, no EOM is generated. The spin transfer for an s-polarized incidence is shown in fig.8b. For an s polarized beam incident on the left side of the slit only plasmons on the left edge of the slit are excited propagating towards the junction carrying a z-component of the spin of $S_z = \frac{h}{2\pi}\cos 45°$. The plasmon is transferred to the right edge where it propagates in the negative y direction. Thus a spin of $-\frac{h}{2\pi}\cos 45°$ is generated which is compensated by an EOM of $L_z = \frac{h}{2\pi}\cos 45°$. For an incidence at the right side the opposite EOM of $L_z = -\frac{h}{2\pi}\cos 45°$ is generated and no EOM is generated for an incidence along the y axis.

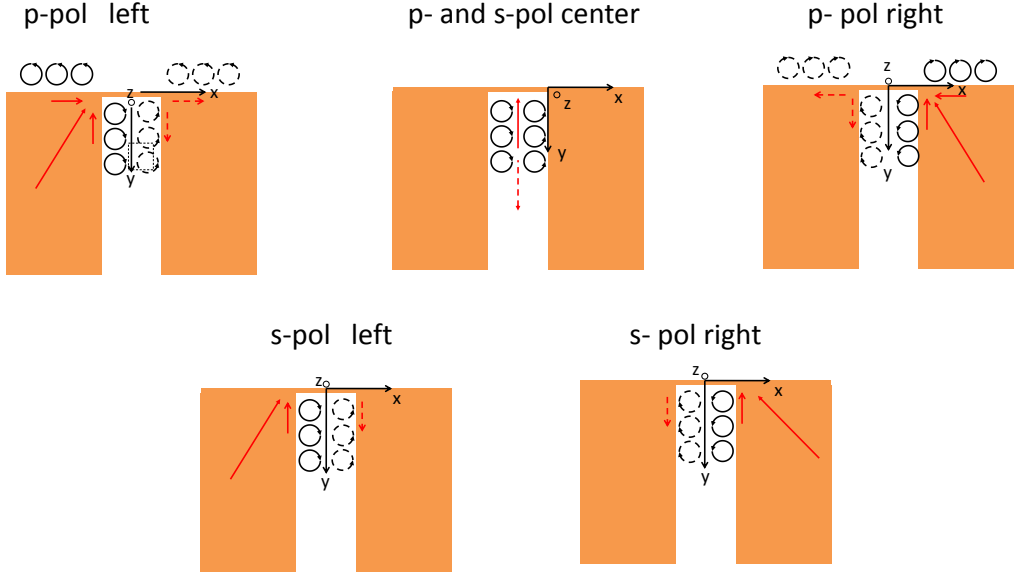

**Fig. 8** *Spin transfer at the slit structure for the scattering process of an incident p- and s-polarized beam of light for different lateral displacements from the slit in the x-direction. The red arrow indicates the direction of the incident beam. The dashed red arrows indicate the direction of the reflected edge plasmon mode. The black solid and dashed circular arrows indicate the turning sense of the z-component of the spin of the edge plasmon and the reflected edge plasmon respectively.*

For p polarization, photons have a polarization in the plane of incidence of the incident beam and they carry a spin in the direction of polarization with a positive z –component $S_z = \frac{h}{2\pi}\cos 45°$. For an incidence of a photon from the left, an edge plasmon mode is excited at the left edge of the slit and at the x edge both propagating towards the junction. They both carry a z-component $S_z = \frac{h}{2\pi}\cos 45°$ of the longitudinal spin of the incident beam. The edge plasmons are transmitted at the constriction to the right edge of the slit and a reflected plasmon with left handed circular polarization propagates in the –y direction. In this process the z spin component is inverted. The change in Spin angular momentum $\Delta S_z = -\frac{h}{\pi}\cos 45°$ at the right edge is compensated by an electron orbital angular momentum $L$. For incidence of the p-polarized photon at the right side of the slit, an electron angular $L_z = -\frac{h}{\pi}\cos 45°$ is generated accordingly. For incidence of a photon along the y axis, no EOM is generated. For an s polarized beam incident on the left side of the slit only plasmons on the left edge of the slit are excited propagating towards the junction carrying a z-component of the spin of $S_z = \frac{h}{2\pi}\cos 45°$. The plasmon is transferred to the right edge where it propagates in the negative y direction. Thus a spin of $-\frac{h}{2\pi}\cos 45°$ is generated which is compensated by an EOM of $L_z = \frac{h}{2\pi}\cos 45°$. For an incidence at the right side the opposite EOM of $L_z = -\frac{h}{2\pi}\cos 45°$ is generated and no EOM is generated for an incidence along the y axis.

For the wedge with normal incidence and the slit we measured the electromotive force as a function of the distance of the axis of the Gaussian beam from the slit or wedge. In both cases the spatial distribution of the signal had a half width of 4 µm. We think this half width is related to the range of the edge Plasmon and edge waveguide modes respectively which are excited by the incident focused beam. Edge plasmons thus may increase the interaction range for generating the spin transfer and the metal edges can be considered as antennas for the transfer of photon spin to electron orbital angular momentum.

For an estimation of the magnitude of the measured voltages we consider the case of normal incidence of a circular polarized beam of numerical aperture 1 into a retroreflecting conducting nanostructure which is not further specified. The spin of the circular polarized photon is inverted and, due to conservation of angular momentum, the momentum is transferred to the retroreflecting structure, such that a circular movement of electrons is generated. If its angular momentum is transferred to electrons it creates an electron motive force corresponding to an electron rotation around a circle of a radius $\lambda/2$ corresponding to what can be regarded as the radius of a photon with an angular momentum $\frac{h}{\pi}$. In this rotation the electron picks up a linear momentum $p$ and a kinetic energy $E_{kin} = \frac{1}{2}\frac{p^2}{m_e}$. With the angular momentum for the movement of the electron around a circle of a radius $\frac{\lambda}{2}$ being equal to $p\frac{\lambda}{2} = \frac{h}{\pi}$ we obtain the relation:

$$V_{emf} = \frac{2\,h^2\nu}{\pi^2\,em_e c\lambda}$$

For the wavelength $\lambda = 655$ nm we obtain the result $V_{emf} = 1{,}416$ µV. This should be the maximum value of any photo effect we can expect in our measurements. It is limited by the energy transfer in the process of the transfer of photon spin to electron orbital angular momentum. Similar as in the case of the photoelectric effect, the electromotive force is not limited by the light intensity i.e. the energy flux but by the kinetic energy which is picked up by an electron in the transfer of photon spin to electron orbital angular momentum. There are different reasons why we were not able to measure the saturation value of the electromotive force.

1) The metal junctions of our structures have a rather low resistance R. If the photoinduced current is smaller than $\frac{V_{Emf}}{R}$, the resistance R acts as a short circuit for the EMF thus decreasing the net light induced current. It should be possible to reach a saturation of the effect by increasing the light intensity of the incident beam. We noticed, however, that increasing this intensity can easily lead to a destruction of the narrow metal junctions. Therefore and because the measured signals were rather small we did not yet systematically investigate the dependence of the signals on light intensity.

2) The spin transfer in our experiments is according to our interpretation less than $\frac{h}{\pi}$ except for the case of the p-polarized beam incident into the wedge at normal incidence.

3) In the derivation of the electromotive force, we assumed a path length of d = $2\pi \frac{\lambda}{2}$ = 2,04 µm for the electron in a circular movement associated with an angular momentum. In the experimental situation the length of the path on which electrons pick up angular momentum may be different and it is not circular. The longer the path, the smaller is the energy loss in the scattering process and the associated kinetic energy which is acquired by the electrons in the process of spin transfer and $V_{emf}$ becomes smaller. We may estimate the path length for the electrons by assuming it to be equal to the half width of 4 µm in our measurements. With this assumption and the estimated spin transfer according to the interpretation of the experiments on the slit samples we fit the measured data of the photo voltage as a function of the beam displacement from the slit. We assume a Gaussian distribution for the stationary density of photons with a half width of 4 µm in the x -direction. We think that the half width of 4 µm is related to the range of the edge plasmon modes. For our fit we use the following semiempirical formula with the parameters :

$$V(x) = cV_{emf} \left\{ e^{-\frac{(x-d)^2}{2D^2}} - e^{-\frac{(x+d)^2}{2D^2}} \right\}$$

In this formula the parameters $c, V_{emf}, d, D$ enter.

$c = 1$ for p-polarisation and $c = 0,5$ for s-polarization, $2d$ signifies the half width of the focus of the incident beam and $D$ signifies the range of edge plasmon modes along the edges of the gold film. The parameters $d = 0,5 µm$ and $D = 4 µm$ were not measured but were derived from fitting the formula to the experimental values of fig.6a. A half width of the incident beam of 1µm and a range of 4 µm for the edge plasmon modes seem to be reasonable parameters for our experimental configuration.

**Conclusions and Outlook**

Partially gold coated 90° glass wedges and a semi - infinite slit in a thin film of gold ending in a conducting junction of a cross section in the order of the skin depth were fabricated as samples in order to investigate the transfer of photon spin to electron orbital angular momentum. These structures were specifically designed as samples where an incident beam of light is retroreflected. In the 90° wedge structure an incident beam is reflected into the reversed direction of incidence. In the slit structure an incident beam excites edge plasmon modes which are excited on one edge of the slit and are reflected at the other edge in the reverse direction. The incident beam is considered as a current of photons. Since in the process of retroreflection the turning sense of a circularly polarized beam of light does not change and the direction of propagation is inverted the photon spin is inverted. Due to conservation of angular momentum a transfer of photon spin to an angular momentum of the retroreflecting structure occurs. Due to the low mass of the electrons, the spin is mainly transferred to electron orbital angular momentum. The time derivative of the electron orbital angular momentum in the process of retroreflection is equivalent to an inertial torque of the photons acting on the conduction electrons of the gold as an electromotive force which leads to the detection of a photovoltage. Depending on the polarization of the incident beam, a maximum photovoltage of about 0,2µV was measured for both structures. The results are interpreted in terms of an electrodynamic model of the monochromatic photon which is elaborated elsewhere. A linearly polarized photon is considered as a solitary electromagnetic wave propagating in the longitudinal direction of the polarization at a speed of $c\sqrt{2}$ with an energy $h\nu$ and a spin angular momentum $\frac{h}{2\pi}$ in the direction of propagation. The frequency is not encoded in an oscillatory stationary vibration of the electromagnetic field of the single cycle solitary wave but in an angular frequency of the spin of the propagating photon. The relative values of the measured photovoltages can well be explained by our electrodynamic model of a photon. The absolute values of the measured photovoltages are also consistent with our interpretation. On the basis of our interpretation of the experiments we predict, that the saturation value of the photovoltage is limited by the transfered photon spin and we do not expect that this value can increase with light intensity. This effect is similar to the photoeffect where a photovoltage is limited by the energy of the photon and not by the intensity. In the case of the spin transfer the photovoltage does, however, not only depend on the spin and energy of the photons but also on the confinement of the focused beam in the scattering object. The incident photon strongly spreads in our scattering object due to the excitation of edge waveguide modes and edge plasmon modes which have a range of several µm. The observed effects are closely related to the lateral Fedorov Imbert shift of focused beams in optics and the optical spin hall effect [16-19] and to other non linear optical effects such as the inverse faraday effect [20] for which a new interpretation is given here in terms of the electrodynamic model of the photon. It should be interesting to devise objects with retroreflective properties which exhibit a strong confinement and enhancement of the electromagnetic fields of an incident beam. In this case, one can expect a reduced Fedorov Imbert shift but strongly enhanced photovoltages.

**Acknowledgements**

This work was performed within a project "Edge plasmon mediated tip enhanced spectroscopy" Fi 608 of the German Science Council DFG. U.C.F and F.F. acknowledge support by short term missions and travel support by the COST Action MP1302 Nanospectroscopy of the European Union.